\documentclass{nature}

\usepackage{acronym}
\usepackage{amsmath}
\usepackage{amssymb}
\usepackage{color}
\usepackage{hyperref}
\usepackage{graphicx}
\usepackage{caption}
\usepackage{subcaption}

\newcommand\msun{M_\odot}
\def\ltsima{$\; \buildrel < \over \sim \;$}
\def\simlt{\lower.5ex\hbox{\ltsima}}
\def\gtsima{$\; \buildrel > \over \sim \;$}
\def\simgt{\lower.5ex\hbox{\gtsima}}

\newcommand{\plotone}[1]{\includegraphics[width=\columnwidth]{#1}}

\newcommand{\aap}{Astron. and Astrophys.}
\newcommand{\apj}{Astrophys. J.}
\newcommand{\apjl}{Astrophys. J. Lett.}
\newcommand{\apjs}{Astrophys. J. Supp.}
\newcommand{\mnras}{Mon. Not. R. Astron. Soc.}
\newcommand{\nat}{Nat.}

\newcommand{\prd}{Phys. Rev. D}
\newcommand{\pasj}{PASJ}

\begin{document}

\acrodef{BBH}{binary black hole}
\acrodef{BH}{black hole}
\acrodef{EM}{electromagnetic}
\acrodef{GW}{gravitational wave}

\title{Linking gravitational waves and X-ray phenomena with joint LISA and Athena observations}

\author{Sean McGee$^{1}$, Alberto Sesana$^{1,2}$, Alberto Vecchio$^{1,2}$}

\maketitle

\begin{affiliations}
\item School of Physics and Astronomy, University of Birmingham,
  Birmingham, B15 2TT, United Kingdom\\
\item Institute for Gravitational Wave Astronomy, University of Birmingham,
  Birmingham, B15 2TT, United Kingdom
\end{affiliations}

\begin{abstract}

The evolution of cosmic structures, the formation and growth of the first black holes and the connection to their baryonic environment are key unsolved problems in astrophysics. The X-ray Athena mission and the gravitational-wave Laser Interferometer Space Antenna (LISA) offer independent and complementary angles on these problems. We show that up to $\sim 10$ black hole binaries in the mass range $\sim$ 10$^5$ - 10$^8$ $\msun$ discovered by LISA at redshift $\lesssim 3.5$ could be detected by Athena in an exposure time up to 100 ks, if prompt X-ray emission of $\sim$ 1\% - 10\% of the Eddington luminosity is present. Likewise, if any LISA-detected extreme mass ratio inspirals occur in accretion disks, Athena can detect associated electromagnetic emission out to redshift $\sim$ 1. Finally, warned by LISA, Athena can point in advance and stare at stellar-mass binary black hole mergers at redshift $\simlt$ 0.1. These science opportunities emphasise the vast discovery space of simultaneous observations from the two observatories, which would be missed if they were operated in different epochs.
\end{abstract}

\acresetall{}

The observation of the gravitational-wave signal GW170817~\cite{2017PhRvL.119p1101A}, the gamma-ray burst GRB170817A~\cite{2017ApJ...848L..13A} and the optical transient AT-2017gfo~\cite{2017Sci...358.1556C} together with emission across the whole electromagnetic spectrum from the region surrounding the binary neutron star merger~\cite{2017ApJ...848L..12A} at the centre of this rather spectacular chain of phenomena has recently shown us the power of multi-messenger and multi-band observations to make major progress in solving long-standing puzzles in astronomy and astrophysics \cite{2017Natur.551...85A,2018Natur.561..355M,2019Sci...363..968G}. Fast-forward fifteen years, and we could witness similar transformational discoveries related to compact objects thousand-to-million times more massive than those observable from the ground, and the associated high-energy transients and their environments. Or may be not.

The Laser Interferometer Space Antenna (LISA~\cite{2017arXiv170200786A}) is an ESA space-based gravitational-wave (GW) observatory, currently planned for launch in 2034. It will survey the gravitational-wave sky in the $10^{-4}$Hz -- 0.1\,Hz frequency band, providing unprecedented capabilities of discovering, and accurately characterising the physics of black holes in binary systems across the mass scale, from stellar objects of $\sim 10\,\msun$, at the low-mass end, to super-massive black holes up to $\sim 10^8\,\msun$ at the other extreme of the mass spectrum. Binary systems composed of compact objects above $\sim 10^3\,\msun$ will be observable throughout the universe, well beyond the epoch of re-ionisation and reaching redshifts $\approx 20$ \cite{2016PhRvD..93b4003K}. LISA is therefore set to drastically change the landscape of massive black holes: those below $\sim 10^6\,\msun$, whose population is poorly known or a total mystery, and those in the mass range $\sim 10^6 - 10^8\,\msun$ where it will discover many more objects, and selected in a very different way, than those observable through surveys in other observational bands, currently in operation or planned. In fact, while electromagnetic probes preferentially select lone, steadily accreting objects \cite{2013arXiv1306.2325A}, LISA will probe highly dynamic, coalescing systems. LISA will therefore provide a wealth of targets to explore black hole demographics, how they grow, pair-up and collide, the environments in which they reside, and to track structure formation through cosmic time using these different tracers. This is a route that may yield some rather spectacular discoveries. As its counterparts on the ground, LISA will however be totally blind to the properties of the hosts of black holes, the traits of baryons and gas surrounding them, and the possible transient high-energy phenomena that accompany the presence and/or merger of these cosmic giants. Understanding how and if massive black holes and galaxies co-evolve, how massive black holes grow their mass, accrete, and the role of circum-binary disks are just a few examples of the rich and currently hotly debated physics that could become accessible. To reconstruct the full and multi-facet aspects of these phenomena, an X-ray telescope capable of deep wide-field imaging and spatially-resolved spectroscopy, which are features of the next generation X-ray telescopes, such as ESA's Athena mission \cite{2013arXiv1306.2307N}, is needed. Combined with optical and radio observations, through e.g. the Large Synoptic Sky Survey (LSST\cite{2009arXiv0912.0201L}) and the Square-Kilometre-Array (SKA\cite{2009IEEEP..97.1482D}), the next decades are well set to provide major advances, and likely big surprises, in the study of these phenomena.

LISA is currently scheduled to be launched after Athena. Overlap of science operations of the two missions would allow simultaneous observations of the same systems in these two radically different observational bands. They would enable opportunities to observe in X-ray GW-selected targets, or to follow up at high energies the site hosting a massive (or a merging pair of massive) black hole(s). Without joint observations, progress would instead rely on understanding these objects and environments independently, and on characterizing their properties through statistical studies of the catalogues of the two missions. Given that missions of the capabilities (and cost) of LISA and Athena do not come along too frequently, an interesting question to consider, which is at the centre of this paper, is: if LISA and Athena were to operate simultaneously, what is the potential of observing the \emph{same} systems, and the associated phenomena in these two bands?  

There are three obvious science targets that could provide unprecedented information about major open questions in astrophysics and cosmology, and open up new avenues for explorations of black hole demographics, accretion physics and the relevant high-energy radiative processes for a potentially large number of new objects: (i) the interplay between gravity and baryons in (super-)massive binary black hole binary (MBHB) mergers along the cosmic history\cite{2003ApJ...582..559V,2004ApJ...611..623S}; (ii) the massive black holes and their environment(s) in the low-redshift universe, revealed by LISA surveys of extreme-mass-ratio inspirals (EMRIs \cite{2004PhRvD..69h2005B,2007CQGra..24R.113A}); and (iii) the environment and possible high-energy phenomena associated with (heavy) stellar-mass binary black holes (BBHs) in the local universe \cite{2016PhRvL.116w1102S}. We will explore them in turn in the following sections. 
 
Investigating whether it is possible to make joint GW-X ray observations of the same environment boils down to answering the following questions: if a black hole binary is detected by LISA, is it feasible to carry out a targeted follow-up X-ray observational campaign with Athena covering the whole GW position error-box on the sky? And, in turn, how deep can the X-ray observations go (and therefore what fraction of GW-detected events could they reach) to detect prompt X-ray emission related to the merger?  The answer to these questions depends on the performances of the instruments at one's disposal and, possibly more significantly, on the actual details of physical processes at work, which currently are largely uncertain, see e.g. \cite{2002ApJ...567L...9A,2005ApJ...622L..93M,2010MNRAS.407.2007C,2010Sci...329..927P,2010ApJ...715.1117B,2014PhRvD..90j4030G,2016MNRAS.457..939C,2018MNRAS.476.2249T,2018ApJ...865..140D}. These uncertainties in turn reflect the substantial discovery space that these observations will access. 

As LISA's reference performance we consider the design described in~\cite{2017arXiv170200786A}: a 2.5 million-km arm interferometer in heliocentric orbit with requirements that lead to the noise spectral density of Fig. 2 of Ref~\cite{2017arXiv170200786A}. For the X-ray observatory we use the design specifications of the Athena WFI detector \cite{2013arXiv1308.6785R}. We assume a field of view of 0.4\,deg$^2$ and divide the frequency band into $[0.5,2]\,$keV (soft, hereinafter) and $[2,10]\,$keV (hard, hereinafter). We assume that the flux sensitivity limit for a $5\sigma$ detection scales with the integration time $T$ as 
\begin{equation}
F_{(X,S)} = 3\times 10^{-16} \left({\frac{10^5\,\mathrm{sec}}{T}}\right)^{1/2} \mathrm{erg}\,\mathrm{cm}^{-2}\,\mathrm{s}^{-1}\,,
\label{eq:Fx_det_lim}
\end{equation}
\begin{equation}
F_{(X,H)}=5F_{(X,S)},
\label{eq:Fx_detH_lim}
\end{equation}
for $1\,$ks$<T<100\,$ks. At $T<1\,$ks observations will likely be photon-limited. Regardless of the brightness of the source, we therefore conservatively take $1\,$ks to be the shortest possible exposure time. At $T>100\,$ks we conservatively assume that the flux limit saturates at $10^{-16}\, \mathrm{erg}\,\mathrm{cm}^{-2}\,\mathrm{s}^{-1}$ in the soft band. Note that this is still a factor of $\approx 4$ larger than the source confusion limit estimated at $2.4\times10^{-17}\, \mathrm{erg}\,\mathrm{cm}^{-2}\,\mathrm{s}^{-1}$ \cite{2013arXiv1308.6785R}.The WFI is assumed to be five times less sensitive in the hard band due to the decline of the effective collecting area at high energies \cite{2013arXiv1308.6785R} (Guainazzi, private communication). For each source, we estimate detectability in the soft and hard bands separately and pick the most favourable one.

\section*{Gravity and baryons in massive black hole binary mergers}
\label{s:MBHB}

LISA detection rate of merging MBHBs is highly uncertain, spanning the range $\sim 10-1000$ for a planned 4 year mission duration~\cite{2016PhRvD..93b4003K}. The mass-redshift distribution of the observed systems is subject to large uncertainties, and based on modelling, the majority of the detections will be of systems with masses $\simlt 10^5\msun$ at redshift $z \simgt 5$, although a few detections of black holes with masses $\simgt 10^6\msun$ at $z\simlt 2$ are expected. This is a quite extraordinary sample of objects, selected in a radically different way from AGN and/or time-domain surveys.

The first step is to identify the performance of LISA in observing MBHBs. For the problem at hand there are two quantitative indicators of the performance: the signal-to-noise ratio (S/N) at which a binary of a given mass and redshift can be observed and the error-box in the sky associated with the detection.  
The specific numbers for a given system depend on a large number of factors that introduce many complications that prevent us from giving a clear-cut picture of the science potential of these observations and go beyond the scope of this work in which we look at broad order of magnitude results. 
In this spirit, to derive an empirical relation between the total mass of the MBHB $M$ and the S/N, $\rho$, at which it is observed, we consider the set of simulations performed by~\cite{2016PhRvD..93b4003K}. They are based on population models from~\cite{2012MNRAS.423.2533B}, which feature different seed black hole formation and subsequent accretion histories, also resulting in a wide range of spin and mass ratio distributions. We include all the observed systems in all models and we find that the median angular resolution ${\Delta{\Omega}}$ is related to the median ${\rho}$ at which a system is observed by 
\begin{equation}
  \Delta{\Omega}\approx 0.5\,\left(\frac{\rho}{10^3}\right)^{-7/4}  {\rm deg}^2\,.
  \label{eq:skyMBHB}
\end{equation}
Note that the relation is slightly flatter than the standard $\Delta \Omega\propto\rho^{-2}$ dependency. This is due to the fact that larger $\rho$ are generally obtained for more massive binaries, which tend to stay in band for a shorter time. The scatter around the median is large, 0.7dex, and is found to be dominated by the range of source sky location, inclination and polarization angles, rather than the individual mass, mass ratios and spin values of the systems.

\begin{figure}
\centering
\begin{minipage}[c]{\textwidth}
\centering
\includegraphics[scale=0.65,clip=true]{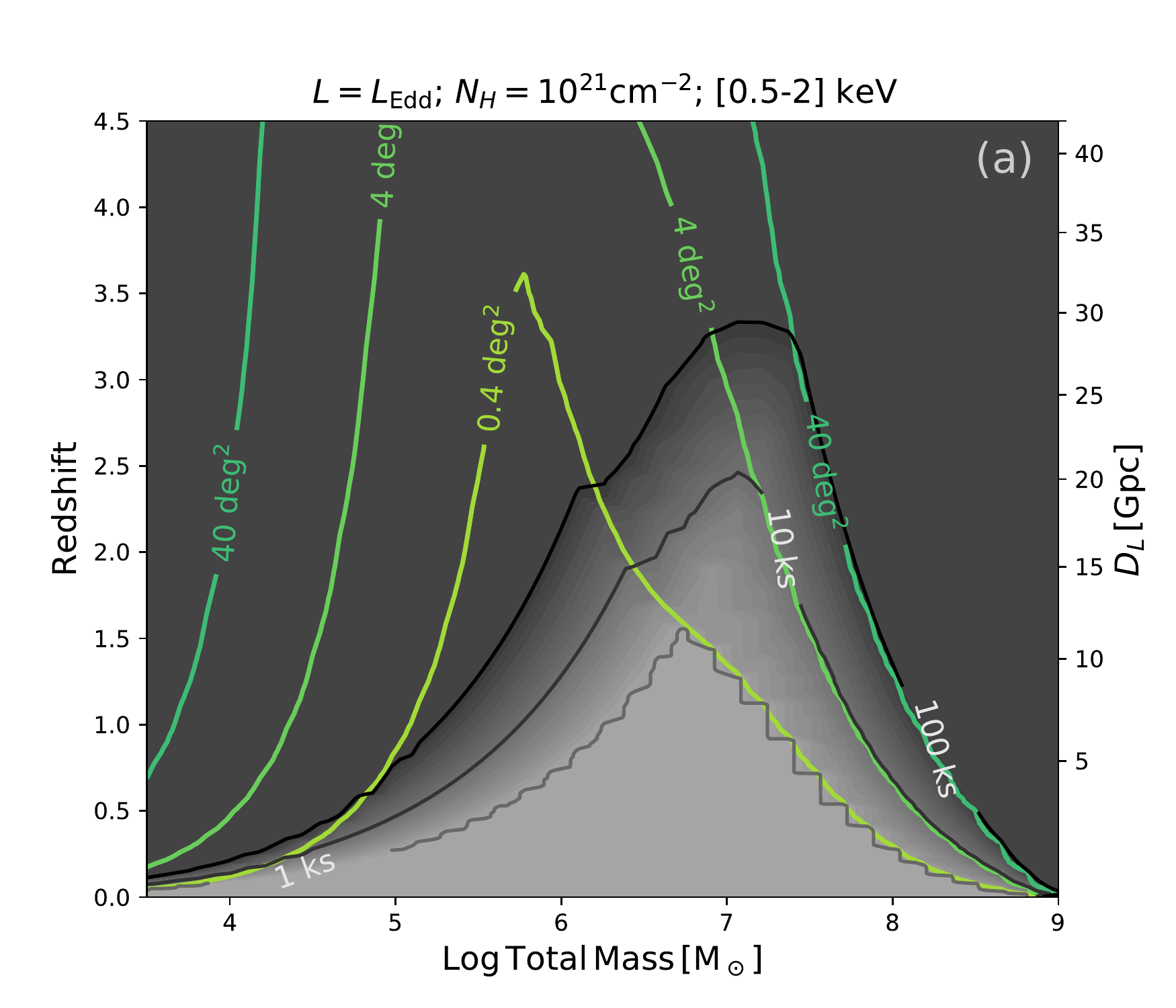}
\end{minipage}
\begin{minipage}[c]{\textwidth}
\centering
\includegraphics[scale=0.4,clip=true]{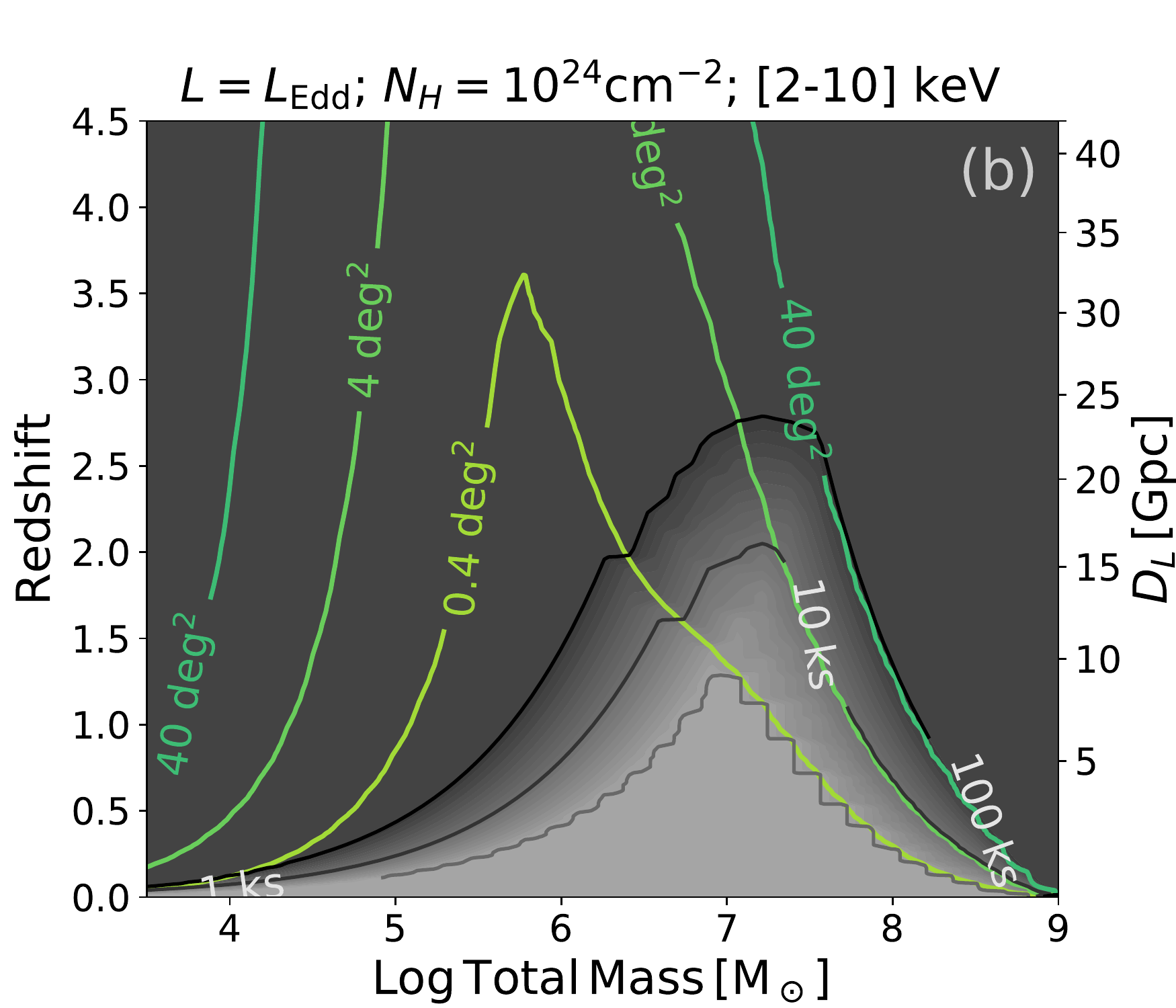}
\includegraphics[scale=0.4,clip=true]{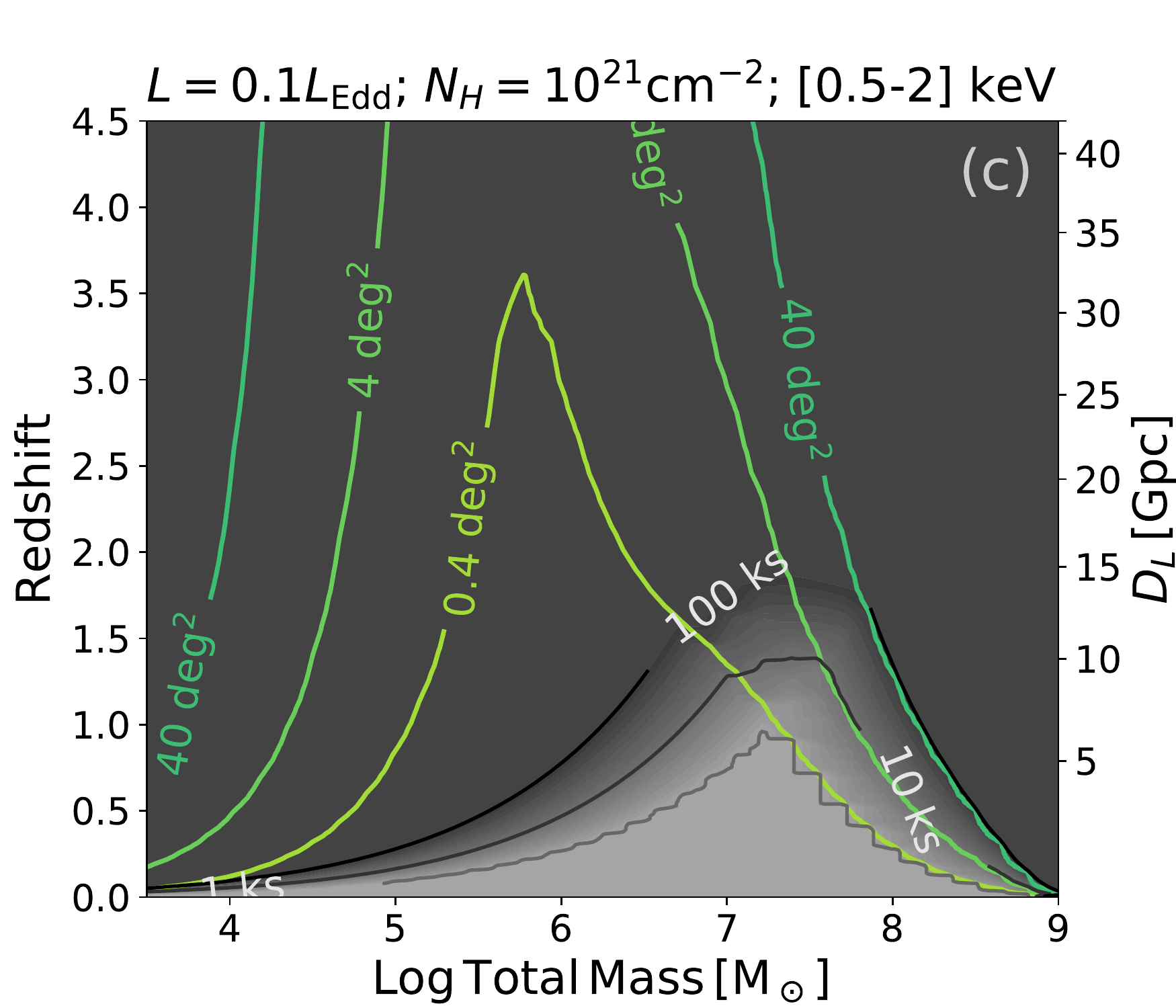}
\vspace{0.5cm}
\end{minipage}
\caption{
   \label{fig:MBHB} 
   \textbf{Feasibility of joint GW/X-ray observations of MBHB mergers in the mass-redshift plane.} In each panel, green-yellow contours mark the median sky location accuracy of  mergers achievable by LISA observations. Grey shaded contours represent the corresponding total exposure time needed by Athena to cover the LISA error-box, while the black lines demarcate the 1, 10, and 100ks exposure time contours. In the top panel (a) we assume a fiducial source bolometric luminosity $L=L_{\rm Edd}$ and negligible obscuration. The bottom panels show the effect of heavy obscuration (b) and dimmer sources emitting at $L=0.1L_{\rm Edd}$ (c). See text for full details.}
\end{figure}

With a scaling in hand, we compute the sky-inclination-polarization averaged S/N for an equal-mass MBHB as a function of total mass $M$ and $z$, using the latest LISA sensitivity curve ~\cite{2017arXiv170200786A} and PhenomC waveforms for non-spinning, circular binaries \cite{2010PhRvD..82f4016S}. To each $(M,z)$ pair we can then associate a median sky localization through Eq.~\eqref{eq:skyMBHB}, which is shown by the green contours in Fig. \ref{fig:MBHB}. Although the scaling was obtained from a binary population featuring more complex properties than the systems used to compute S/N's on the mass--redshift grid, we note that the distribution of MBHB mass ratios in ~\cite{2016PhRvD..93b4003K} peaks at unity; moreover the exact spin value does not have a major impact on the S/N calculation which makes the use of non spinning waveforms acceptable.

We can now turn to the question of whether any X-ray radiation could be observed. In the absence of solid predictions, in this feasibility study we conservatively use a standard quasar template to model the spectral energy distribution of a merging MBHB. The X-ray portion of the spectrum is modelled as a single power law with spectral index $\alpha=0.7$ consistent with the average quasar emission \cite{2000MNRAS.316..234R}, normalized so that 3\% of the bolometric luminosity goes into the hard (i.e., $[2,10]\,$keV) band \cite{2012MNRAS.425..623L}. In our fiducial scenario we take the  Eddington luminosity, $L_\mathrm{Edd} = 1.26 \times 10^{38}\,\left({M}/{\msun}\right)\,\mathrm{erg}\,\mathrm{s}^{-1}$ as a proxy for the bolometric luminosity of the source, but we also  consider the case of dimmer sources, shining at a level of $0.1L_\mathrm{Edd}$. To account for complications arising from less favourable observational conditions, we also investigate the effect of obscuration due to photoelectric absorption up to an equivalent hydrogen column density of $N_H=10^{24}\,\mathrm{cm}^{-2}$. For all cases, we make the assumption that every merging MBHB is associated with an AGN. Mergers are known to trigger copious gas infall into the center of the merger remnant resulting in a local gas rich-environment and significant AGN accretion \cite{1996ApJ...471..115B, 2008ApJS..175..356H}. In all environments, only about 1\% (10\%) of MBHs show accrection at $\sim$ 1 (0.1) $L_{\rm Edd}$ \cite{2012MNRAS.425..623L, 2013MNRAS.428..421S}. If merging MBHBs do not preferentially occur in gas rich environments our predictions might be as much as $\sim$ 10 times too high.

For a $L_{\rm Edd}$ source modelled as described above, the emission in the $[0.5,10]\,$keV band is $L_X\approx1\times 10^{43}\,(M/10^6\,\msun)\,$erg s$^{-1}$. For any LISA source mass and redshift, we can then compute fluxes in the soft and hard bands. The integration time required to achieve a $5\sigma$ detection in a single pointing in either band is then set by Eq.~(\ref{eq:Fx_det_lim}) and ~(\ref{eq:Fx_detH_lim}). Finally we obtain the required total observations time by combining the number of pointings needed to cover the GW error box, i.e. we multiply by $\Delta\Omega/0.4\,\mathrm{deg}^2$ (being $0.4\,\mathrm{deg}^2$ the WFI field of view). In our analysis we therefore implicitly assume that the emission is persistent at a roughly constant level over the Athena's observation time, which might be the case if there is enough gas in the vicinity of the binary to light a quasar following its final coalescence, as found in some recent simulations, e.g.~\cite{2018MNRAS.476.2249T}. The integration time is shown in the $(M,z)$ plane by the grey-colour map in all panels of Fig. \ref{fig:MBHB}. 

Panel (a) of Fig. \ref{fig:MBHB} shows the fiducial case of $L=L_\mathrm{Edd}$ and no significant obscuration ($N_H=10^{21}\,\mathrm{cm}^{-2}$). Under these assumptions, detection is more efficient in the soft band. Interestingly, Athena has the potential to detect a counterpart of a typical $\sim 10^6-10^7\msun$ merging binary radiating at Eddington out to $z\approx 1.5$ in a single pointing of $1\,$ks duration. Note that for such masses, the dynamical timescale at merger is of the order of $1\,$ks as well, implying that a bright X-ray transient of similar duration might be plausible. Such transient could be prompted by ``gas squeezing" in which case it might be highly super-Eddington \cite{2002ApJ...567L...9A,2016MNRAS.457..939C}, facilitating detection by Athena at even larger $z$.  Reaching out to $z\approx3.5$ would generally require several pointings for a total time of up to 100 ks. Panel (b) shows the case of heavy obscuration, $N_H=10^{24}\,\mathrm{cm}^{-2}$. Under these conditions, detection is still possible in the hard band out to $z\lesssim 3$, with only moderate loss compared to the unobscured case. Finally, the effect of intrinsically dimmer sources (but no significant obscuration) is shown in panel (c), where $L=0.1L_\mathrm{Edd}$ is assumed. In this latter case there is a significant loss, with sources observable in X-rays only out to $z\lesssim 2$.

Although our understanding of the MBHB merger history is poor, it is nevertheless useful to check what is the number of systems expected in the region jointly probed by LISA and Athena. For this exercise we proceeded as follows. We took the three population models discussed in Ref \cite{2016PhRvD..93b4003K}. For each model we computed the distribution $d^2N/dMdz$ observed by LISA in four years. We then integrated the portion of distribution lying below the 100 ks contour shown in Figure \ref{fig:MBHB} for each separate scenario, which defines those systems that can be detected by Athena within a total integration time of 100 ks. Depending on the MBHB population model, for unobscured $L=L_\mathrm{Edd}$ sources (panel (a) in Fig. \ref{fig:MBHB}) we obtain between 3 and 8 LISA-Athena joint detections. Between 2 and 4 sources are detectable in the case of heavy obscuration (panel (b) in Fig. \ref{fig:MBHB}). And only $\lesssim 1$ if the intrinsic luminosity is  $L=0.1L_{\mathrm{Edd}}$ (panel (c) in Fig. \ref{fig:MBHB}). Note that this calculation is approximate because it makes use of the median LISA sky localization at each $(M,z)$. We verified that detected numbers change by less then 20\% when a realistic sky localization distribution is considered at each $(M,z)$. Relaxing our assumption that all MBHBs are associated with an AGN, we conclude that four years of joint LISA-Athena operations can potentially result in 0.1-to-10 joint detections of merging MBHBs.

The scientific potential of joint GW and X-ray detections is vast. Detecting an X-ray flare at merger or afterglow will inform us on the (thermo)dynamics of accretion disks in the strong, dynamic field of a MBHB. For example, a super Eddington flare is indicative of dynamical gas squeezing, which can be related to the thermodynamic and geometry of the minidisks around the MBHs just before merger \cite{2016MNRAS.457..939C}. Post merger bursts due to shocks developing in the disk can be related to GW recoils \cite{2010MNRAS.401.2021R}. In the case of a progressive source re-brightening we will be witnessing the birth of an AGN in real time. We will then be in the unprecedented position to study the onset of accretion onto an object of known mass and spin. Detection of jet launching might allow to establish whether there is a direct connection with black hole spin or not \cite{1977MNRAS.179..433B}.

\section*{The low-mass end of the massive black hole mass function}

\begin{figure}
  \plotone{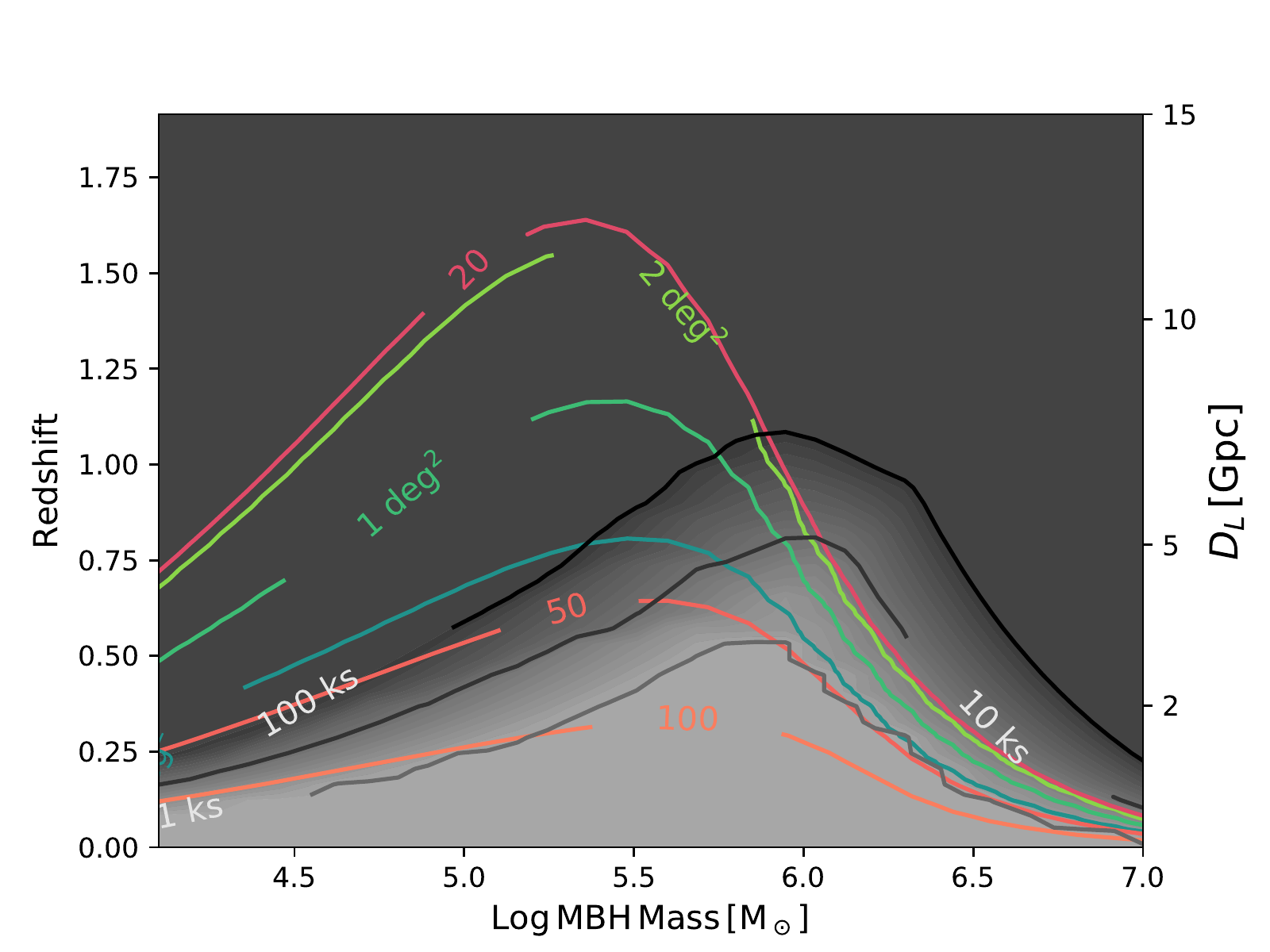}
  \caption{
   \label{fig:EMRI}
   \textbf{Feasibility of joint GW/X-ray observations of a stellar-mass compact object (here assumed to be a $10\,\msun$ BH) spiralling into a massive companion (EMRI) in the MBH mass-redshift plane}. Red contours indicate the S/N of LISA detections. Green-yellow contours mark the associated size of the median error box in the sky containing the GW source. Grey shaded regions are the exposure time for Athena observations to detect an MBH accreting at Eddington, while the black lines demarcate the 1, 10, and 100ks exposure time contours.
   }
\end{figure}

LISA will also map the population of MBHs at low redshift (up to $z\approx 2$) through the detection of low-mass compact objects (neutron stars and stellar-mass black holes) spiralling into them, the so-called extreme mass ratio inspirals (EMRIs)\cite{2007CQGra..24R.113A}. Also in this case, the event rate is highly uncertain, ranging from just a handful to possibly thousands during the mission lifetime \cite{2017PhRvD..95j3012B}. Most of these systems will be characterised by MBH masses of $\simlt 10^6\msun$, therefore probing a region of the MBH parameter space very poorly known \cite{2010PhRvD..81j4014G}. Note that standard EMRI formation channels are not related to MBH accretion, see for a review~\cite{2007CQGra..24R.113A}, which will allow us to access a sizeable sample of the completely unexplored low mass quiescent MBH population, inaccessible, by definition, to electromagnetic observations. 

Similarly to the MBHB case, we consider the depth of a GW observation as a function of the MBH mass, $M$ (which is essentially the total mass of the system given the extreme mass ratio), in the $(M, z)$ plane. We set the compact object companion to have a mass of $10\msun$.  We then relate the typical GW sky error box to the detection S/N by considering the extensive set of simulations reported in~\cite{2017PhRvD..95j3012B}, for which we obtain the empirical fit:
\begin{equation}
  \Delta{\Omega}\approx 0.05\,\left(\frac{\rho}{100}\right)^{-5/2}  {\rm deg}^2,
  \label{eq:skyEMRI}
\end{equation}
which relates the median sky resolution $\Delta{\Omega}$ to the median S/N $\rho$.  Here the slope is steeper than $\rho^{-2}$ (although the scatter, 0.4dex, is still large).
As for the MBHB case, we take a grid of points in $(M,z)$, compute sky- and inclination-averaged $\rho$ using AK waveforms assuming a plunge eccentricity of $e_p=0.2$ consistent with Ref.~\cite{2017PhRvD..95j3012B} and assign a median $\Delta{\Omega}$ based on Eq. \eqref{eq:skyEMRI}. The results of this procedure are shown in Fig. \ref{fig:EMRI} by the colour contours. There are two obvious differences with respect to the MBHBs case: events are observed only up to moderate redshifts $z \simlt 2$,
and the associated $\Delta{\Omega}$ is much smaller, of the order of $1\,\mathrm{deg}^2$ or smaller. The largest uncertainty region shown corresponds to $\Delta\Omega= 2$\,deg$^2$ which is the typical sky localization accuracy of a LISA threshold detection (assumed at $\rho=20$, see red contours in Fig. \ref{fig:EMRI}). 

Although EMRI formation is generally not associated with AGN activity, but see for an alternative scenario~\cite{2007MNRAS.374..515L}, a fraction of EMRIs will nonetheless occur in AGNs. This is particularly interesting since the drag from the accretion disk might imprint an observable signature in the EMRI dynamics, thus informing us that the inspiral is occurring around an accreting MBH \cite{2011PhRvL.107q1103Y,2014PhRvD..89j4059B}. The identification of a counterpart AGN will allow us to connect the drag force to the properties of the accretion disk (spectrum, accretion rate, variability, emission lines, etc) thus providing invaluable insights on the nature of the disk viscosity, which is currently poorly understood. Moreover, the evolution of the X ray luminosity and of the shape of the K$\alpha$ line (if detected) as the stellar BH spirals in \cite{2013MNRAS.432.1468M}, might be used to constrain the size and geometry of the hot corona and the reflection properties of the accretion disk, adding important elements for the development of a full theory of accretion physics.

We thus evaluate what Athena may be able to reveal by using for the accreting MBH the same quasar template adopted in the previous section. Estimates of AGN activity in low mass black holes are highly uncertain. Only $\sim$ 1\% of the expected host galaxies have optical signatures of AGN\cite{2013ApJ...775..116R}, but such measures are only sensitive to rapidly accreting black holes within galaxies of low star formation rates. If the AGN rate is very low, a LISA pre-selection of EMRI targets with evidence for an environmental effect on the waveform may be necessary. Assuming an EMRI is identified in an AGN, then we further assume emission at the Eddington level and negligible obscuration. The total exposure needed to cover the LISA error-box is shown by the grey colour scale in Fig. \ref{fig:EMRI}. Note that, under these assumptions, most of the sources at $z<0.5$ would be detected at the inexpensive cost of $\approx 1$ ks. Therefore, exposures extending to $\sim 100$ ks would allow us to detect MBHs accreting at a mere $\sim$ 1\% Eddington. Moreover, for a few percent of the detected EMRIs, the 3-dimensional error box from GW observations will be sufficiently small to contain just one Milky Way-like galaxy, allowing a much more systematic observational campaign with other facilities. 

\section*{Stellar mass binary black holes}

\begin{figure}
  \plotone{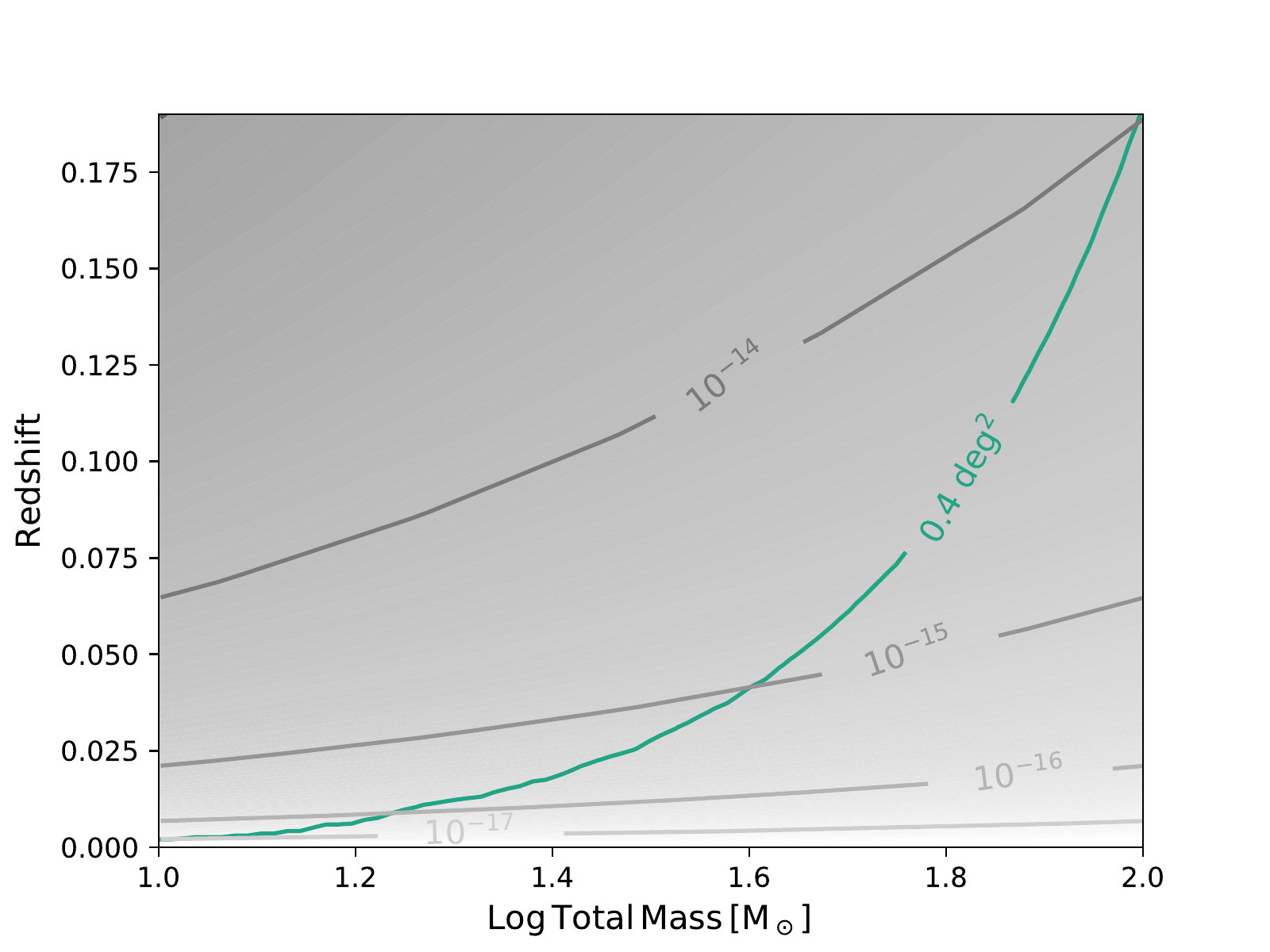}
  \caption{
   \label{fig:SOBH}
   \textbf{The discovery space in the total mass-redshift plane of joint GW/X-ray observations of stellar mass binary black holes.} The green line marks the median LISA sky localisation accuracy corresponding to $0.4$\,deg$^2$, which corresponds to the Athena field of view, so that systems below this line can be followed-up in X-ray with a single pointing. Grey shaded contours represent the 5$\sigma$ limit that an Athena non-detection will impose on the mass-energy conversion efficiency into X-ray (see text for details), while the black lines demarcate the 10$^{-16}$, 10$^{-15}$ and 10$^{-14}$ erg s$^{-1}$ cm$^{-2}$ contours.
   }
\end{figure}

LISA will be able to observe a consistent population of inspiralling stellar-mass binary black holes (BBHs) with $M>30\msun$ at $z<0.5$~\cite{2016PhRvL.116w1102S}. A subset of them, will be caught only few years before final coalescence, an event that will then be observed by ground-based GW observatories. For those systems LISA will determine the time to merger with a statistical uncertainty of $\approx 1$ sec, and the sky location with a precision smaller than the Athena WFI field of view, several weeks before the final coalescence takes place. In fact, taking the results of~\cite{2018MNRAS.475.3485D} the median sky-localisation is 
\begin{equation}
  \Delta{\Omega}\approx 0.2\,\left(\frac{\rho}{10}\right)^{-2}  {\rm deg}^2.
  \label{eq:skySOBH}
\end{equation}
This will allow us to point Athena in the right place at the right time, and stare at the site of a collision of BBHs when this catastrophic event takes place. Although progress is continuously being made for ground-based technology, and third-generation GW ground-based observatories, e.g.~\cite{2010CQGra..27s4002P, 2017CQGra..34d4001A}  are expected to be in operation together with a space-based mission, it is unclear whether they will able to provide sufficiently prompt advanced warning to trigger X-ray observations, simply due to the short time these events are in band, $\approx 15.5 \eta (f_\mathrm{low}/3\,\mathrm{Hz})^{-8/3}\,(M/20\msun)^{-5/3}$ min, where $\eta$ is the symmetric mass ratio and $f_\mathrm{low}$ the low-frequency cut-off of the instrument's sensitivity window. 

If a merging BBH were to accrete at the Eddington limit, it would produce a flux in X-rays of $\approx 10^{-17} (M/10\,\msun)\,(D_L/1\,\mathrm{Gpc})^{-2}$ erg s$^{-1}$ cm$^{-2}$. Limits for e.g. GW150914 ($D_L \approx 400$ Mpc) have been set to an X-ray flux of $\approx 1-0.1\times 10^{-9}$ in the 2-20 keV band ~\cite{2017PASJ...69...84K}. A number of hypothetical scenarios have been put forward to produce bright supernovae precursors \cite{2018ApJ...855L..12M}, prompt GRB emission \cite{2016ApJ...821L..18P} and luminous afterglows \cite{2017ApJ...839L...7D} of BBH mergers. In all cases, the emission would be strong enough that it would be easily observable by Athena. Detection of bright emission from these objects will revolutionize our understanding of their formation, dynamics and environment.

Here, we take the standard assumption that the merger occurs approximately in vacuum, but consider the possibility of an exotic, yet unknown, mechanisms able to convert a fraction $\epsilon$ of the total binary mass-energy into radiation during the formation of the remnant black hole. We can then calculate what is the minimum mass-to-radiation conversion efficiency $\epsilon_{\rm th}$ that can be probed by Athena. We assume, for simplicity, that all the electro-magnetic energy is released in X-rays, and that photons are emitted with energy $E_\gamma$ uniformly distributed in the energy range $\Delta E_\gamma \sim E_\gamma$. The associated number of photons is thus $N_\gamma=\epsilon Mc^2 /E_\gamma$, which gives 
\begin{equation}
  N_\gamma\approx 10^{64}\epsilon\left(\frac{M}{10\msun}\right)\,\left(\frac{E_\gamma}{1\,\mathrm{keV}}\right)^{-1}.
\end{equation}
Assuming isotropic emission, a source at a luminosity distance $D_L$ and a collecting area of a wide field instrument $A_\mathrm{WFI}$, the number of photons impinging the X-ray detector is
\begin{equation}
N_{\rm WFI} = 10^{15}\epsilon\left(\frac{M}{10\msun}\right)\left(\frac{A_\mathrm{WFI}}{1\,\mathrm{m}^2}\right)\,\left(\frac{D_L}{300{\rm Mpc}}\right)^{-2}\, {\rm cts}. 
\end{equation}
If the burst is produced by some exotic mechanism associated to the formation of the remnant black hole, it will occur on the merger timescale, $\tau \sim 10^{-4}(M/\msun)$ s. The number of photons collected during the burst has to be compared to the background level of the instrument integrated over the time $\tau$. The irreducible background limit of Athena is expected to be at the level of
$F_{\rm bkg}\approx 5\times10^{-3} {\rm cts}\,\, {\rm cm}^{-2} {\rm s}^{-1} {\rm keV}^{-1}$,
which is irrelevant for a point source with typical spot size on the focal plane of $\approx0.03$ cm (Piro, private communication). The requirement for a positive detection is therefore determined by the minimum number of photons incident on the detector, which we set to 10. The minimum conversion efficiency probed by Athena is simply given by
 \begin{equation}
  \epsilon_{\rm th}=\frac{10}{N_{\rm WFI}}=10^{-14}\left(\frac{M}{10\msun}\right)^{-1}\left(\frac{A_\mathrm{WFI}}{1\,\mathrm{m}^2}\right)^{-1}\left(\frac{D_L}{300{\rm Mpc}}\right)^{2}. 
 \end{equation}
 This is shown by the shaded grey scale in Fig. \ref{fig:SOBH}, where we have assumed $A_\mathrm{WFI}=1$ m.

\section*{Conclusions and caveats}
 
The results presented here are affected by very considerable (by orders of magnitude) uncertainties in the underlying physical processes. We also caution that detection is different then identification. One major challenge for Athena will be to identify the true GW event counterpart among hundreds of candidates in the field of view. Assessing the feasibility of this task requires deeper investigations and extensive simulations of Athena mock observations. This is beyond the scope of this Perspective article, but is a crucial bottleneck to maximise the science return of these observations. If on the one hand the possibility of joint GW-X-ray observations should be approached with considerable caution, on the other hand this highlights the tremendous opportunity of (possibly multiple) discoveries and transformational observations.

\begin{addendum}
\item SMG acknowledges the support by UK's Science and Technology Facilities Council (STFC); AS is supported by a University Research Fellowship of the Royal Society; AV acknowledges support from STFC, UK Space Agency, the Royal Society and the Wolfson Foundation. We thank E. Barausse for providing the MBHB population models discussed in the text. 
\item[Data Availablity] The source data used in the figures is available on request from the corresponding author.
\item[Author Contributions] All authors contributed to the work
  presented in this paper.
\end{addendum}

\bibliographystyle{naturemag}

\end{document}